\documentclass[twocolumn,showpacs,superscriptaddress,PRB]{revtex4}

\usepackage{amsmath}
\usepackage{amssymb}
\usepackage{graphicx}

\begin{document}

\title{Scanning Tunneling Microscopy/Spectroscopy of Vortices in LiFeAs}

\author{T. Hanaguri}
\affiliation{RIKEN Advanced Science Institute, Wako, Saitama 351-0198, Japan}
\affiliation{TRiP, Japan Science and Technology Agency, Chiyoda, Tokyo 102-0075, Japan}

\author{K. Kitagawa}
\altaffiliation[Present address: ]{Graduate School of Integrated Arts and Sciences, Kochi University, Kochi 780-8520, Japan}
\affiliation{Institute for Solid State Physics, University of Tokyo, Kashiwa, Chiba 277-8581 Japan}

\author{K. Matsubayashi}
\affiliation{TRiP, Japan Science and Technology Agency, Chiyoda, Tokyo 102-0075, Japan}
\affiliation{Institute for Solid State Physics, University of Tokyo, Kashiwa, Chiba 277-8581 Japan}

\author{Y. Mazaki}
\affiliation{Institute for Solid State Physics, University of Tokyo, Kashiwa, Chiba 277-8581 Japan}
\affiliation{Department of Physics, Nihon University, Chiyoda, Tokyo 101-8308, Japan}

\author{Y. Uwatoko}
\affiliation{TRiP, Japan Science and Technology Agency, Chiyoda, Tokyo 102-0075, Japan}
\affiliation{Institute for Solid State Physics, University of Tokyo, Kashiwa, Chiba 277-8581 Japan}

\author{H. Takagi}
\affiliation{RIKEN Advanced Science Institute, Wako, Saitama 351-0198, Japan}
\affiliation{TRiP, Japan Science and Technology Agency, Chiyoda, Tokyo 102-0075, Japan}
\affiliation{Department of Physics, University of Tokyo, Bunkyo, Tokyo 113-0033 Japan}

\date{\today}

\begin{abstract}
We investigate vortices in LiFeAs using scanning tunneling microscopy/spectroscopy.
Zero-field tunneling spectra show two superconducting gaps without detectable spectral weight near the Fermi energy, evidencing fully-gapped multi-band superconductivity.
We image vortices in a wide field range from 0.1~T to 11~T by mapping the tunneling conductance at the Fermi energy.
A quasi-hexagonal vortex lattice at low field contains domain boundaries which consist of alternating vortices with unusual coordination numbers of 5 and 7.
With increasing field, the domain boundaries become ill-defined, resulting in a uniformly disordered vortex matter.
Tunneling spectra taken at the vortex center are characterized by a sharp peak just below the Fermi energy, apparently violating particle-hole symmetry.
The image of each vortex shows energy-dependent 4-fold anisotropy which may be associated with the anisotropy of the Fermi surface.
The vortex radius shrinks with decreasing temperature and becomes smaller than the coherence length estimated from the upper critical field.
This is direct evidence of the Kramer-Pesch effect expected in a clean superconductor.
\end{abstract}


\maketitle

\section{Introduction}

Vortices in type-II superconductors have attracted much attention.
Since superconductivity is destroyed at the vortex center, the vortex core can be regarded as a potential well surrounded by the superconducting gap (SG).
Therefore, quasi-particle states in the vortex core provide us with useful information of the SG.
In this respect, it is particularly interesting to study vortices in iron-based superconductors~\cite{Kamihara2008JACS,Paglione2010NatPhys,Hirschfeld2011ROPP,Stewart2011RMP}.
Characteristic features of iron-based superconductors, namely high transition temperature, multi-band nature,~\cite{Singh2008PRL} small carrier density and possible sign-reversal in the SG,~\cite{Mazin2008PRL,Kuroki2008PRL,Seo2008PRL,Wang2009PRL,Chubukov2009PRB,Cvetkovic2009EPL,Christianson2008Nature,Chen2010NatPhys,Hanaguri2010Science} should bring about unusual quasi-particle states in and around the vortex core.
Here we investigate vortices of an iron-based superconductor using scanning tunneling microscopy/spectroscopy (STM/STS).

In general, the excitation spectrum in the vortex core consists of discrete bound states separated by $\sim\Delta^2/E_F$, where $\Delta$ is the SG and $E_F$ is the Fermi energy.~\cite{Caroli1964PhysLett}
In conventional superconductors, $E_F$ is many orders of magnitude larger than $\Delta$ and the excitation spectrum can be regarded as continuous.
Depending on the quasi-particle lifetime $\tau$ associated with impurity scattering, the vortex core can be classified into either dirty or clean regimes.
If the lifetime broadening $\hbar/\tau$ is greater than $\Delta$ (equivalent to $\xi>l$, where $\xi$ is the coherence length and $l$ is the mean-free path), quasi-particles are frequently scattered in the core and the core is well approximated as a normal metal.
The influence of SG on the core states is masked in this dirty case.
On the other hand, if $\hbar/\tau<\Delta$ or $\xi<l$, quasi-particles can travel from one side of the vortex core to the other without being scattered.
The most prominent feature of such a clean vortex core is that local density-of-states (LDOS) at the vortex center exhibits a peak at $E_F$ because low-energy bound states are localized near the center.
This LDOS peak was first observed by Hess {\it et al.} in NbSe$_2$~\cite{Hess1990PRL} and subsequent theoretical analysis revealed that the energy and spatial evolution of the LDOS peak reflect the nature of SG as well as normal-state band structure~\cite{Gygi1990PRL,Gygi1991PRB,Hayashi1996PRL,Hayashi1997PRB}.

If the ratio $E_F/\Delta$, which roughly corresponds to the number of bound states, becomes small, energy separation between the bound states is large and the excitation spectrum is no longer continuous, provided $\hbar/\tau<\Delta^2/E_F$.
Unfortunately, this so-called quantum-limit regime may not be reachable in conventional phonon-mediated superconductors.
However, in iron-based superconductors, because of small carrier density (which leads small $E_F$) and high transition temperature (which means large $\Delta$), the quantum-limit regime may be anticipated.
Theoretical analysis of the quantum-limit vortex predicts strong particle-hole asymmetry in an LDOS spectrum at the vortex center and a Friedel-like oscillation in the spatial variation of LDOS.~\cite{Hayashi1998PRL}

So far, vortices are successfully observed by STM/STS in some iron-based supercondcutors.~\cite{Yin2009PRL,Shan2011NatPhys,Song2011Science}
The LDOS peak associated with the bound states has been detected in the tunneling spectra at the vortex centers of FeSe and Ba$_{0.6}$K$_{0.4}$Fe$_2$As$_2$.~\cite{Song2011Science,Shan2011NatPhys}
In the latter compound, possible quantum-limit behavior is discussed based on the asymmetry of the spectrum.~\cite{Shan2011NatPhys}
However, detailed electronic structure of the vortex core and effects of multi band are still elusive.

In addition to the quasi-particle states in the vortex core, properties of many vortices, vortex matter, have provided a fertile area of research.
In general, vortices tend to form a triangular lattice (Abrikosov lattice) due to the repulsive interaction between them.
However, the Abrikosov lattice may qualitatively be altered by quenched disorders and/or thermal agitation.~\cite{Blatter1994RMP}
The competition among associated energy scales, namely, elastic energy of the Abrikosov lattice, pinning energies of quenched disorders and thermal energy, brings about various vortex phases: hexatic Abrikosov-like phase (elastic energy dominant), vortex glass or amorphous vortices (pinning energy dominant) and vortex liquid (thermal energy dominant).
The effect of thermal agitation and consequent vortex-lattice melting transition has extensively been studied in cuprate high-temperature superconductors.
However, competition between elastic and pinning energies is still to be clarified.

The stability of the Abrikosov lattice is determined by its shear modulus $C_{66}$ which is roughly given by $B_{c2}^2b(1-b)^2$ where $B_{c2}$ is the upper critical field and $b$ is the magnetic field $B$ normalized by $B_{c2}$.~\cite{Brandt1986PRB}
At low and high $b$ where $C_{66}$ is small, the Abrikosov lattice should be destroyed by the quenched disorders and the amorphous vortex matter may emerge.
The vortex lattice partially retains its order in the intermediate $b$ range but should contain dislocations.~\cite{Larkin1979JLTP}
Numerical simulation of the vortex configuration shows that vortices with unusual coordination numbers of 5 and 7 are aligned alternately, forming domain boundaries.~\cite{Chandran2004PRB}
A crossover to the high-field amorphous phase occurs through roughening of the boundaries.

In this article, we report results of comprehensive STM/STS experiments on vortices in an iron-based superconductor LiFeAs.
LiFeAs is one of the very few stoichiometric iron-based superconductors and thus clean single crystals are available.
In addition to this, cleaved surfaces studied by STM/STS are electronically neutral, which is highly advantageous to minimize the surface-related extrinsic effects.
Tunneling spectrum exhibits clear two-gap features indicating that the SG amplitude is different depending on the Fermi-surface cylinder.
In contrast to other vortex-imaging techniques such as Bitter decoration, STM/STS can image vortices even at high magnetic field where $B$ is almost uniform.
By taking this advantage, we study the structure of vortex matter over wide $B$ range and find a signature of high-field crossover to the amorphous phase.
We also examined the electronic state of the vortex cores.
Tunneling spectra at the vortex center are strongly particle-hole asymmetric with a sharp LDOS peak just below $E_F$.
We compare the spectra and their spatial evolution with those expected in the quantum-limit vortex core and show that the LDOS features associated with the larger SG are consistent with the quantum-limit behavior, but those for the smaller SG including the sharp peak just below $E_F$ are not compatible with the quantum-limit vortex-core state.
We also found that vortex radius shrinks with decreasing temperature.
This is a direct evidence of so-called Kramer-Pesch effect~\cite{Kramer1974ZPhys,Hayashi2005JLTP} associated with the low-energy bound states in the vortex core.

\section{Experimental method}

LiFeAs single crystals were grown by a LiAs self-flux technique.~\cite{Imai2011JPSJ}
A resistivity $\rho$ ratio $\rho\rm{(300~K)}/\rho\rm{(20~K)}$ of as high as 45 and no detectable residual specific heat coefficient are observed.
These data demonstrate the high quality of the crystals.
Superconducting transition temperature $T_c$ defined at zero-resistance is about 18~K.
Since LiFeAs degrades in an ambient atmosphere, all of the sample handling procedures were carried out inside of glove boxes filled with purified Ar or N$_2$ gases.
All STM/STS experiments were performed in a constant-current mode using a commercial ultra-high-vacuum low-temperature STM system modified by ourselves (Unisoku USM-1300 with Nanonis controller).~\cite{Hanaguri2006JPhys}
Tunneling conductance, which is proportional to the LDOS at given bias voltage, was measured by standard lock-in technique using the built-in software lock-in detector of Nanonis controller.
Clean and flat surfaces perpendicular to the [001] axis were prepared by {\it in situ} cleaving at liquid N$_2$ temperature.
Immediately after the cleaving, samples were transferred to the STM unit which was precooled below 10~K.
Electrochemically etched W wires, which were cleaned, sharpened and characterized by a field-ion microscope, were used for scanning tips.
Magnetic fields were applied along the [001] axis.
Whenever magnetic field was changed, the sample was heated up above $T_c$ in advance and then it was field-cooled to the measurement temperatures in order to ensure the homogeneous field distribution inside the sample.

\section{Results and discussion}

\begin{figure*}
\includegraphics[width=160mm]{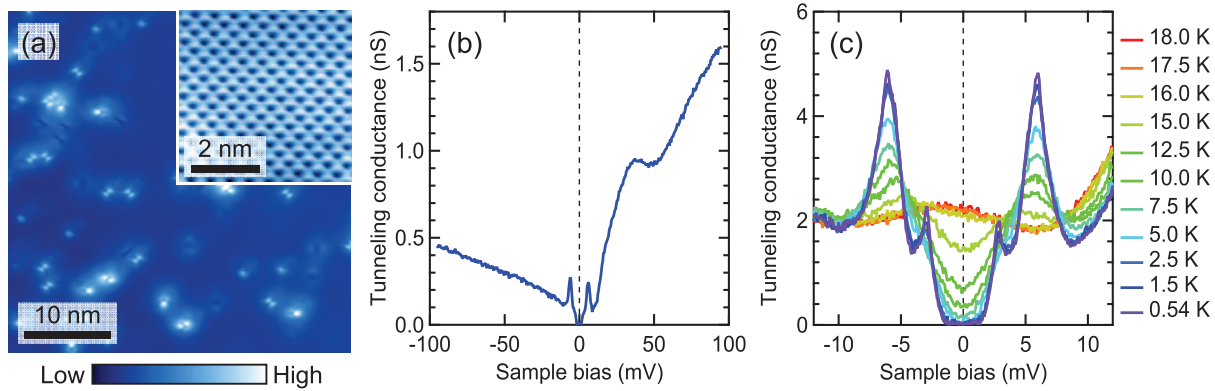}
\caption{(color online).
(a)
STM topograph of cleaved (001) surface of LiFeAs at 0.54~K.
The set-up conditions for imaging were sample-bias voltage $V_s=+20$~mV and tunneling current $I_t=10$~pA.
The inset shows a magnified topography with atomic resolution ($V_s=+20$~mV, $I_t=100$~pA).
(b)
Tunneling spectrum taken away from defects.
Tip stabilization condition was $V_s=+98.9$~mV and $I_t=100$~pA.
Measurement was done with a bias modulation amplitude $V_{\rm mod}=1$mV$_{\rm rms}$.
Data were taken at 1.5~K.
(c)
Low-energy tunneling spectra at different temperatures.
Two gaps are identified in the spectra and both gaps disappear above 18~K which corresponds to bulk $T_c$.
Tip was stabilized at $V_s=+25$~mV and $I_t=125$~pA.
$V_{\rm mod}=0.1$~mV$_{\rm rms}$.
}
\end{figure*}

\subsection{Topography and spectra in zero magnetic field}

We show a typical STM topography in Fig.~1(a).
Although various types of natural defects exist, most of the cleaved surface is clean and flat enough to image a regular square atomic lattice as shown in the inset.
The lattice constant is determined to be 0.39~nm, indicating that the imaged atomic site is either the As site or Li site.

Figure~1(b) represents the tunneling spectrum measured in the clean region.
The spectrum is skewed V-shaped bottomed at $E_F$ with a bump at about 40~meV.
We confirmed that the spectrum does not exhibit obvious spatial variations except in the vicinity of defects.
Results of spectroscopic studies near defects will be published everywhere.

At low energy, SG structure is clearly identified.
As shown in Fig.~1(c), spectral weight over a finite energy range near $E_F$ is completely removed at low temperatures.
This is a direct evidence that the SG is finite everywhere on the Fermi surface.
Two pairs of peaks can be clearly identified, implying that the SG consists of at least two components.
By defining the SG amplitude $2\Delta$ using the peak-to-peak separation, the SG ratio $2\Delta/k_BT_c$ ($k_B$ denotes Boltzmann constant) for the smaller SG ($\Delta\sim2.9$~meV) is estimated to be 3.7, which is close to the value expected from the weak-coupling Bardeen-Cooper-Schrieffer theory.
The SG ratio for the larger SG ($\Delta\sim6.0$~meV) takes a very large value of 7.7.
In both SGs, the rising edges at lower energy have finite width which is wider than our energy resolution ($\sim0.1$~meV).
This result is consistent with the anisotropic SG suggested by recent angle-resolved photoemission spectroscopy (ARPES) experiments~\cite{Umezawa2012PRL,Borisenko2012Symmetry,Okazaki_PC} and quasi-particle interference effect.~\cite{Davis_PC}

The observed SG amplitudes are quantitatively consistent with the ARPES data.~\cite{Umezawa2012PRL,Borisenko2012Symmetry,Okazaki_PC}
We identify that the smaller SG opens on the outer-most hole cylinder and the larger SG is associated with the inner hole cylinders.
A contribution from the electron cylinders, where ARPES observed $\Delta\sim4$~meV,~\cite{Umezawa2012PRL,Borisenko2012Symmetry} is not clear in the tunneling spectrum.
Possible reason for this apparent absence may be that the gap-edge peak for the electron cylinders overlaps with those for the hole cylinders, or electron cylinders are "masked" by the tunneling matrix element.

At elevated temperatures, both gaps are smeared out and disappear at about bulk $T_c$ of 18~K, indicating that they faithfully represent bulk superconductivity.

\subsection{Imaging vortex matter}

Since superconductivity is destroyed at the vortex center, LDOS below the SG energy should be enhanced around the vortices.
Therefore, vortices can be imaged by spectroscopic imaging technique in which tunneling conductance is measured at every pixel of STM image.
Figures 2(a)-(c) are conductance maps at $E_F$ at representative $B$'s, showing vortices as bright spots.
We repeated the same experiment over wide $B$ range from 0.1~T to 11~T.
As shown in Fig.~3(a), $B$ dependence of vortex density perfectly follows the trend expected in the case that single vortex carries the single flux quantum of 2.07$\times10^{-15}$~Wb.
\begin{figure*}
\includegraphics[width=150mm]{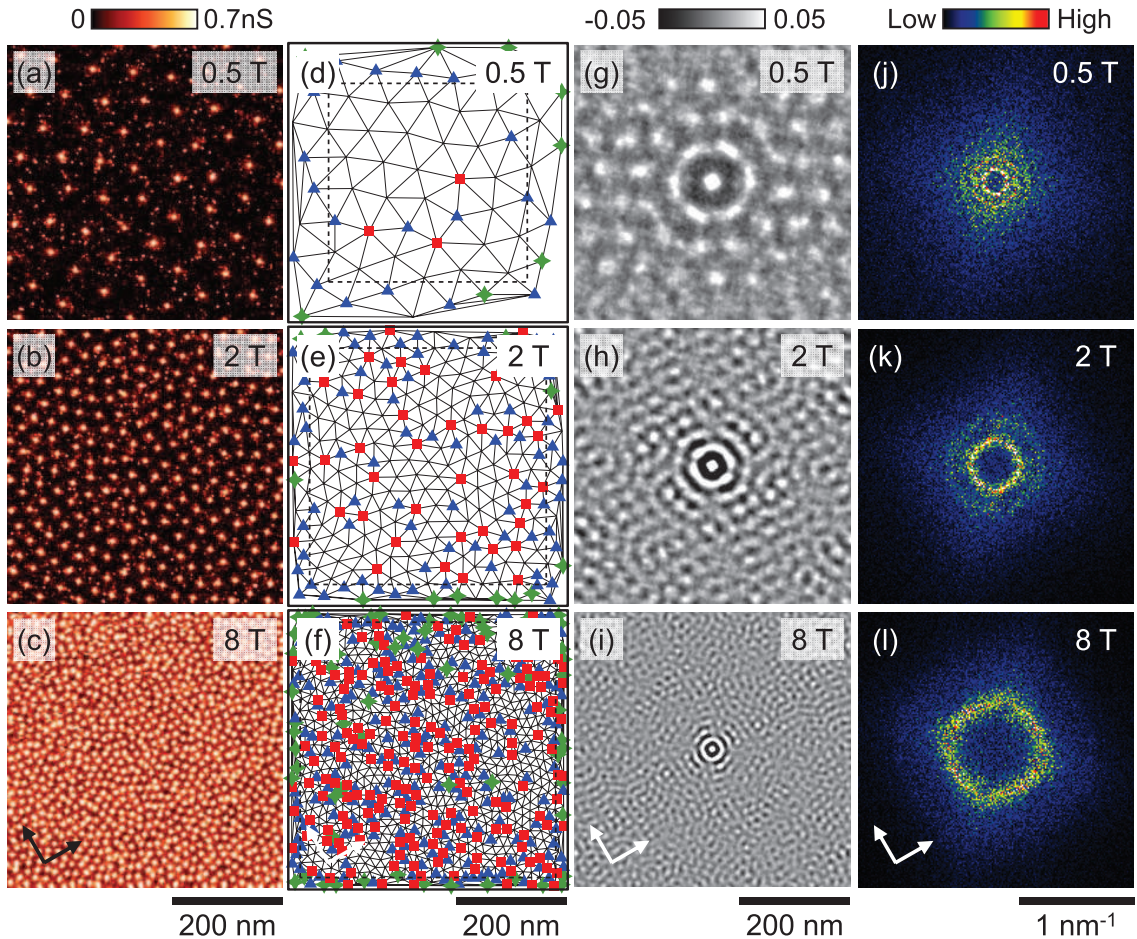}
\caption{(color online).
(a)-(c)
Images of vortices at 1.5~K obtained by mapping tunneling conductance at $E_F$.
Tip was stabilized at $V_s=+20$~mV and $I_t=100$~pA.
$V_{\rm mod}=0.7$~mV$_{\rm rms}$.
Arrows denote the nearest Fe-Fe direction.
(d)-(f)
Delaunay triangulation diagrams obtained from vortex images shown in (a)-(c).
Vertices with symbols denote vortices with coordination numbers different from conventional value of 6. (Blue triangle: 5, red square: 7, green star: others)
Dashed black square is drawn separately from the edge of the field of view (black solid square) by the mean value of inter-vortex distance at each field.
Defect fraction (Fig.~3(b)) was calculated using vortices within this dashed black square to avoid the effect of the edge.
(g)-(i)
Two-dimensional autocorrelation-function calculated from vortex images shown in (a)-(c).
(j)-(l)
Two-dimensional Fourier transformations from vortex images shown in (a)-(c).
}
\end{figure*}

Interestingly, vortices do not form into the ordered triangular lattice.
Two-dimensional auto-correlation (Figs.~2(g)-(i)) and Fourier-transform (Figs.~2(j)-(l)) images are characterized by $B$-dependent ring-like structures around the origin.
At 0.5~T, weak 6-fold spots are identified along the ring in the auto-correlation image (Fig.~2(g)), indicating that hexatic correlation is developing.
At higher $B$, the ring-like structures are isotropic in magnitude but the rings in the Fourier-transform images are deformed into rectangular shape elongated along the As-As direction.
The emergence of 4-fold anisotropy suggests that the symmetry of underlying crystal lattice and/or anisotropy of the SG affect the configuration of vortices but they are not strong enough to arrange vortices into regular square lattice which was observed in YNi$_2$B$_2$C, for example.~\cite{Sakata2000PRL,Nishimori2004JPSJ}.
Nevertheless, the observed ring-like structures indicate that there is a well-developed distance correlation between vortices but the orientational correlation is almost absent.

Such a highly disordered vortex matter in LiFeAs is in sharp contrast to the vortices in Ba$_{0.6}$K$_{0.4}$Fe$_2$As$_2$ where well-ordered triangular vortex lattice has been observed.~\cite{Shan2011NatPhys}
The difference between these two materials can be ascribed to the difference in the shear modulus $C_{66}$ associated with the difference in $B_{c2}$.~\cite{Brandt1986PRB}
In LiFeAs, $B_{c2}$ along the [001] axis is about 15~T,~\cite{Kurita2011JPSJ} while that of (Ba,K)Fe$_2$As$_2$ is estimated to be more than 60~T.~\cite{Altarawneh2008PRB,Yuan2009Nature}
This difference gives rise to an order-of-magnitude smaller $C_{66}$ in LiFeAs than that in (Ba,K)Fe$_2$As$_2$.
Therefore, vortex lattice in LiFeAs is much more susceptible to quenched disorders than that in (Ba,K)Fe$_2$As$_2$.
In other words, LiFeAs can be a playground to investigate how the structure of vortex matter subject to quenched disorder changes as a function of $B$.

In order to analyze the local structure of vortex matter, we performed Delaunay triangulation analysis as summarized in Figs.~2(d)-(f).
Each vertex denotes the position of vortex and vertices with symbols correspond to defect vortices whose coordination numbers are different from 6.
We plot defect-vortex density as a function of $B$ in Fig.~3(b).
In the low field region, defect density increases rapidly and reaches more than 40 \%.

At $B=0.5$~T, a domain boundary, which consists of alternating defect vortices with coordination numbers of 5 and 7, has been observed.~(Fig. 2(d))
More grain boundaries are observed at 2~T but each grain boundary still keeps its alternating structure.~(Fig. 2(e))
At 8~T, defects tend to cluster and defect vortices with extraordinary coordination numbers of 4 and 8 emerge.~(Fig. 2(f))
In order to examine the defect structure within the cluster at 8T, we compare local vortex configurations at 2~T and 8~T in the fields of view which contain equal number of vortices.~(Figs.~3(c),(d))
In contrast to the domain boundary at 2~T, well-defined alternating line-like structure is no longer observed at 8T.
Namely, even though defect density saturates above $\sim$2~T, local vortex configuration still changes with $B$.
\begin{figure}
\includegraphics[width=85mm]{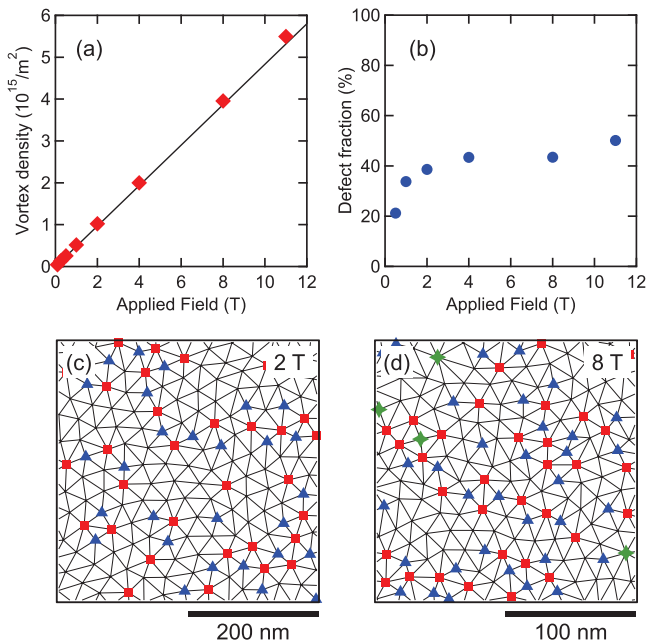}
\caption{(color online).
(a)
Applied-field dependence of vortex density in LiFeAs.
Solid black line denotes the behavior expected in the case that single vortex carries the single vortex quantum of 2.07$\times10^{-15}$~Tm$^2$.
(b)
Applied-field dependence of fraction of defect vortices with coordination numbers different from conventional value of 6.
(c) and (d)
Delaunay triangulation diagrams at 2~T and 8~T in the fields of view which contain equal number of vortices.
Symbols have the same meanings as in Figs.~2(d)-(f).
}
\end{figure}

These observations suggest the following picture of order-disorder crossover in the vortex matter.
In the $B$ region where the elastic energy dominates, number of defect vortices is small.
Even if there is a defect, its coordination number should be 5 or 7, which is close to the regular value of 6.
Such defect vortices are aligned alternately forming a domain boundary.
This situation does not change so much as long as the elastic energy is relevant.
At high enough $B$, however, $C_{66}$ diminishes.
As a result, the domain boundary becomes rough and the vortex matter approaches to the uniformly disordered amorphous phase which contains vortices with extraordinary coordination numbers less than 5 and more than 7.
These behaviors are totally consistent with the result of numerical simulation.~\cite{Chandran2004PRB}

\subsection{Shape of a single vortex}

We now focus on spectroscopic features of a single vortex.
All of the data shown below are taken at $B=$~0.5~T and the reproducibility of the results has been checked in a different tip-sample combination.
Figures 4(a)-(d) show conductance maps at different energies around a vortex.
Vortex-core state at $E_F$ exhibits a 4-fold star shape with high LDOS tails along the nearest As-As directions.
Each tail splits in parallel with increasing energy.
These behaviors closely resemble the energy-dependent vortex-core shape of NbSe$_2$,~\cite{Hess1990PRL} except that in NbSe$_2$ vortex core has a 6-fold anisotropy rather than 4-fold.

Theoretical models exist which account for the energy-dependent vortex-core shape include effects of vortex lattice~\cite{Gygi1990PRL,Gygi1991PRB}, SG anisotropy~\cite{Hayashi1996PRL,Hayashi1997PRB} and crystal-lattice potential which determines the shape of the Fermi surface.~\cite{Gygi1990PRL,Gygi1991PRB,Zhu1995PRB} 
In the case of LiFeAs, the effect of vortex lattice can be safely excluded because vortices at 0.5~T form into a distorted hexatic lattice (Fig.~2(a)), which does not fit with the observed 4-fold anisotropy of the vortex core.

At low enough energies, LDOS distribution around a vortex is primarily governed by the smallest SG and the Fermi surface where it opens.
In LiFeAs, ARPES observed the smallest SG in the outer-most hole cylinder and the SG exhibits 4-fold anisotropy.~\cite{Umezawa2012PRL,Borisenko2012Symmetry,Okazaki_PC}
The SG is smaller along the Fe-Fe direction than along the As-As direction, which would give rise to the vortex-core shape with the LDOS tails along the Fe-Fe direction.
This is in contradiction to the observed LDOS distribution at $E_F$ shown in Fig.~4(a), indicating that the SG anisotropy does not play a major role.

On the other hand, the shape of the Fermi surface is consistent with the observed anisotropy of the vortex-core state.
According to ARPES, the outer-most hole cylinder has a rounded square cross section with flat region perpendicular to the As-As direction.~\cite{Borisenko2010PRL}
This anisotropy should result in a larger number of quasi-particles traveling along the As-As direction, forming high LDOS tails around a vortex.
Indeed, Wang, Hirschfeld and Vekhter analyzed the LDOS distribution around a single vortex in LiFeAs with the assumptions of various SG anisotropy and concluded that Fermi-surface anisotropy in LiFeAs is strong enough to dominate the various types of SG anisotropies.~\cite{Wang2011PRB}
\begin{figure}
\includegraphics[width=80mm]{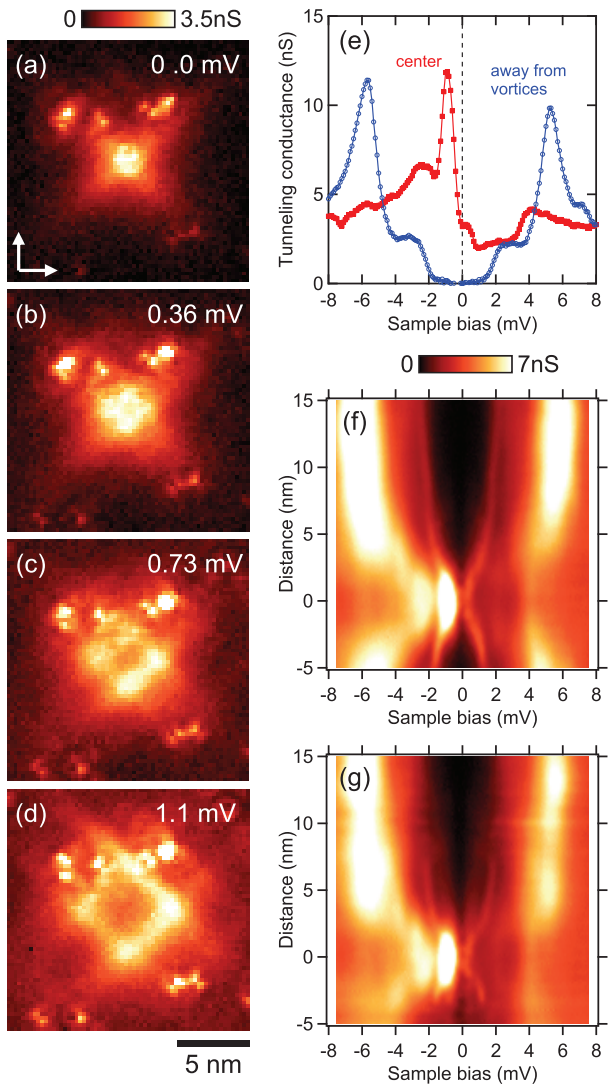}
\caption{(color online).
(a)-(d)
Tunneling-conductance images around a single vortex at different energies.
Arrows denote the nearest Fe-Fe direction.
Tip was stabilized at $V_s=+25$~mV and $I_t=250$~pA.
$V_{\rm mod}=0.2$~mV$_{\rm rms}$.
Data were taken at 0.54~K.
(e)
Tunneling spectra taken at the center of vortex (red) and away from vortices (blue).
$V_{\rm mod}=0.10$~mV$_{\rm rms}$.
Line profiles of tunneling conductance across the vortex center.
(f) Along the nearest Fe-Fe direction.
(g) Along the nearest As-As direction.
$V_{\rm mod}=0.15$~mV$_{\rm rms}$.
}
\end{figure}

\subsection{LDOS spectrum and its spatial evolution}

In order to get more insight into the vortex core state, we study the individual tunneling spectrum and its detailed spatial evolution.
Figure~4(e) shows the spectrum taken at the vortex center.
We also plot the spectrum away from vortices where the two-gap feature observed in the absence of $B$ is well maintained.
The spectrum at the center exhibits a pronounced peak just below $E_F$ (-0.9~meV) as well as some other weaker peaks.
False color plots of spatial evolutions of tunneling spectrum along Fe-Fe and As-As directions are shown in Fig.~4(f) and (g), respectively.
The LDOS peak at $\sim$-0.9~mV shifts further away from $E_F$ with increasing distance from the vortex center and smoothly approaches the smaller SG.
The counter branch is also observed at symmetric energies above $E_F$, although the intensity is much weaker near the vortex center.
There is an energy-gap-like feature apparent at $E_F$ even at the vortex center.
Spectral weight at low energies is higher along the As-As direction than along the Fe-Fe direction, giving rise to the anisotropic shape of the vortex.

LDOS features associated with the larger SG show more complicated behavior.
A broad LDOS peak at $\sim$-2.3~mV at the vortex center exhibits jumps in its energy with increasing distance from the center and finally develops as the peak at the larger-SG energy.
Above $E_F$, the gap-edge peak evolves from the peak at $\sim$+4.2~mV at the center.
Namely, the quasi-particle branches associated with the larger SG are asymmetric between filled and empty states not only in their intensity but also in their energy.
These features are in sharp contrast to the smooth particle-hole symmetric evolutions observed for the quasi-particle branches associated with the smaller SG.

The observed vortex core states in LiFeAs are apparently different from those of conventional superconductors, e.g., NbSe$_2$.
In NbSe$_2$ where the vortex-core excitation spectrum in almost continuous, LDOS peak is located just at $E_F$.~\cite{Hess1990PRL}
With increasing distance from the vortex center, the peak splits along smooth branches which are symmetric in both intensity and energy between filled and empty states.~\cite{Hess1990PRL}
By contrast, in LiFeAs, spectral weight in the vicinity of vortex center is strongly particle-hole asymmetric, there is a gap-like feature even at the vortex center, and evolution of quasi-particle branches for larger SG is neither smooth nor symmetric.
These observations indicate that factors which are lacking in the vortex of conventional superconductor play a vital role.

Proximity to the quantum limit is a plausible candidate of the observed asymmetry, as has been suggested in Ba$_{0.6}$K$_{0.4}$Fe$_2$As$_2$.~\cite{Shan2011NatPhys}
In the quantum-limit vortex, theoretical calculation of LDOS spectrum predicts strong particle-hole asymmetry at the vortex center which is associated with the lowest bound state.~\cite{Hayashi1998PRL}
LDOS peak for the lowest bound state appears only on empty (filled) state if the carriers are electron (hole).~\cite{Hayashi1998PRL,Shan2011NatPhys}
This is apparently consistent with the observed asymmetry in LiFeAs; the LDOS peaks are located below $E_F$ and the relevant SGs are opening in the hole cylinders.

In this scenario, energy of the LDOS peak $|E_p|$ at the vortex center represents the energy of the lowest bound state given by $\Delta^2/2E_F$.
For the larger SG, we observed that $E_p\sim-2.3$~meV and $\Delta\sim6.0$~meV, leading $E_F\sim7.8$~meV.
This $E_F$ value is very small, in general sense, but does not contradict with the ARPES results~\cite{Borisenko2010PRL,Umezawa2012PRL,Borisenko2012Symmetry,Okazaki_PC} where the top of the inner hole band is very close to $E_F$.
The particle-hole asymmetric spatial evolution of the LDOS peak, especially the step-like change in its energy in the filled state, may be due to the Friedel-like oscillation.~\cite{Hayashi1998PRL}
Namely, it is possible that the observed asymmetric quasi-particle states associated with the larger SG reflect the quantum nature of the vortex electronic states in LiFeAs.

However, the quantum-limit picture may not apply for the smaller SG because calculated $E_F\sim4.7$~meV from $E_p\sim-0.9$~meV and $\Delta\sim2.9$~meV is much (at least order-of-magnitude) smaller than the value expected from ARPES dispersion.~\cite{Borisenko2010PRL,Umezawa2012PRL,Borisenko2012Symmetry,Okazaki_PC}
In addition, the quasi-particle branch approaching to the smaller SG is observed to be very smooth without any trace of Friedel-like oscillation.
Such a smooth evolution may suggest that quantum nature is absent in the vortex electronic states associated with the smaller SG.
It is an important open issue to identify what gives rise to the pronounced particle-hole asymmetry in the spectral weight and opens an energy gap at the vortex center.

\subsection{Temperature dependence: Kramer-Pesch effect}

Next we examine temperature $T$ dependence of the vortex-core states.
Figure~5(a) shows the tunneling spectra taken at the vortex center at different $T$.
The LDOS peak at -0.9~meV diminishes rapidly with increasing $T$.
We compare the observed spectrum at each $T$ with the spectrum at 0.54~K convoluted with the Fermi-Dirac distribution function at $T$.
It is clear that the observed LDOS peak is broader than that expected from the thermal smearing effect.
This is in stark contrast to the $T$ dependence of the spectrum away from vortices where each spectrum is well reproduced by the convoluted spectrum as shown in Fig.~5(b).
This contrasting behavior strongly suggests that the temperature not only smears the quasi-particle states but also changes the excitation spectrum in the vortex core.
\begin{figure}
\includegraphics[width=85mm]{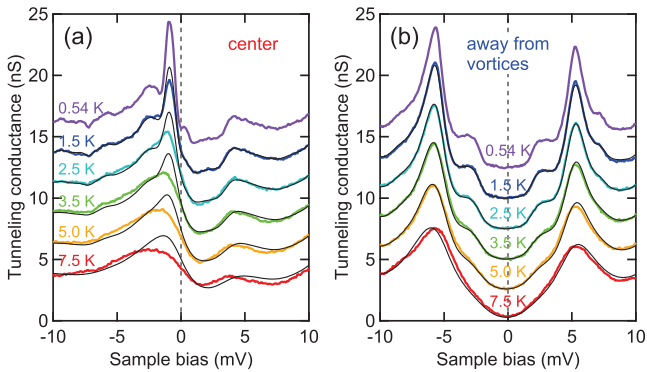}
\caption{(color online).
Temperature dependence of the tunneling spectra at the vortex center (a) and away from vortices (b).
Tip was stabilized at $V_s=+25$~mV and $I_t=250$~pA.
$V_{\rm mod}=0.1$~mV$_{\rm rms}$.
Each curve (except at 7.5~K) is shifted vertically for clarity.
Thin black lines denote the convoluted 0.54~K spectra by Fermi-Dirac distribution function.
}
\end{figure}

The most plausible origin of this unusual temperature effect is the so-called Kramer-Pesch (KP) effect~\cite{Kramer1974ZPhys,Hayashi2005JLTP}.
In clean superconductors where $\hbar/\tau<\Delta$, quasi-particle distribution below $\Delta$ depends on $T$.
The lower-energy bound states are more and more populated at lower $T$.
This $T$-dependent quasi-particle distribution gives rise to the $T$ dependence of spatial variation of SG around a vortex, and vice versa.
Kramer and Pesch first analyzed this process using quasi-classical Eilenberger equation and pointed out that, with lowering $T$, the spatial variation of SG deviates from $\sim\tanh(r/\xi)$ behavior expected from Ginzburg-Landau theory.~\cite{Kramer1974ZPhys}
Here, $r$ denotes distance from the vortex center.
In this case, it is convenient to introduce a new length scale $\xi_1$ defined as $\xi_1=\lim_{r \to 0} (r/\Delta(r))\Delta(r\rightarrow\infty)$ to measure the vortex-core size.
Kramer and Pesch found that $\xi_1$ shrinks linear in $T$, provided $k_BT>\Delta^2/E_F$~\cite{Kramer1974ZPhys}.
Later on, Gygi and Schl{\" u}ter analyzed a clean vortex core based on the quantum Bogoliubov-de Gennes formalism.
As well as reproducing the shrinkage of vortex core, they clarified the $T$ dependence of vortex bound states.~\cite{Gygi1991PRB}

For the sake of examining the relevance of the KP effect, the analysis of the $T$ dependence of tunneling spectrum would not be appropriate because individual bound states are not resolved in Fig.~5(a).
Instead, the $T$ dependence of the vortex-core size can be a direct evidence of the KP effect.
Figure~6 shows the conductance maps at $E_F$ taken at various $T$.
\begin{figure}
\includegraphics[width=70mm]{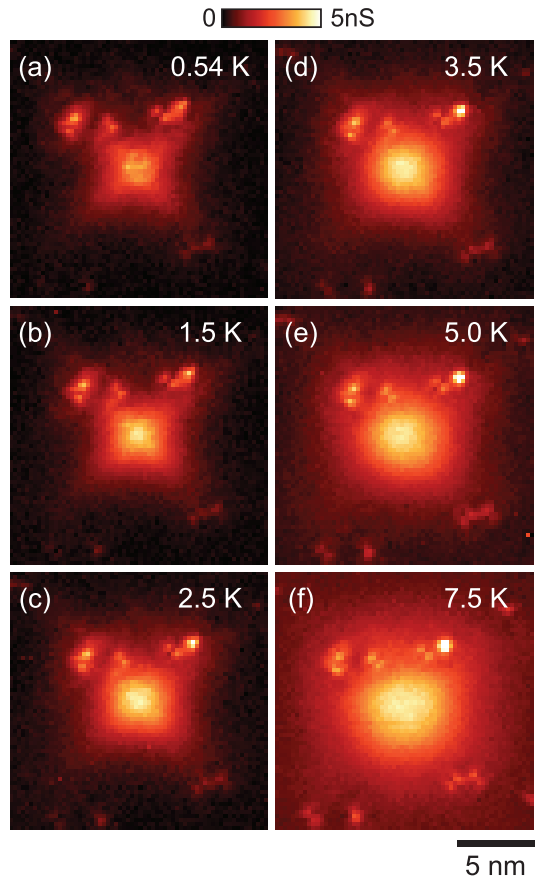}
\caption{(color online).
Tunneling-conductance images at $E_F$ at different temperatures.
Tip was stabilized at $V_s=+25$~mV and $I_t=250$~pA.
$V_{\rm mod}=0.2$~mV$_{\rm rms}$.
Note that color scale is different from that for Fig.4(a)-(d).
}
\end{figure}
It is clear that apparent vortex size increases with increasing $T$, as expected from the KP effect.
In order to quantify this process, we take conductance profiles along Fe-Fe and As-As directions at each $T$.~(Figs.~7(a) and 7(b))
\begin{figure*}
\includegraphics[width=140mm]{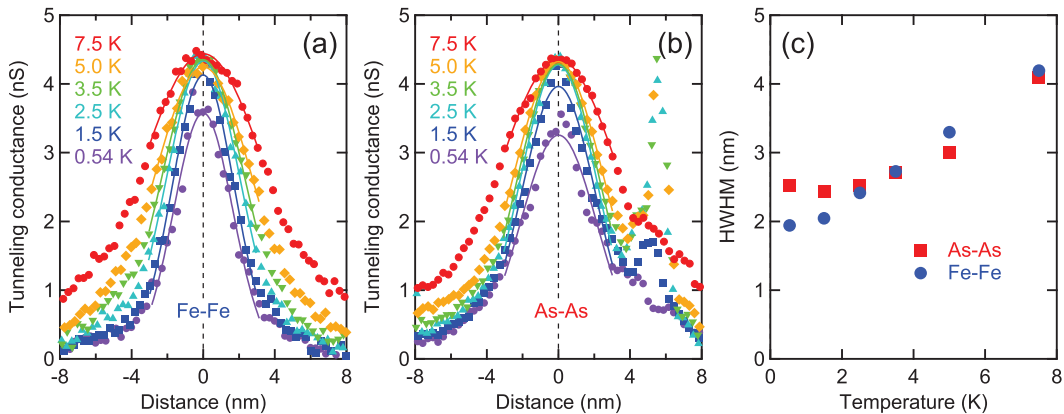}
\caption{(color online).
Line profiles of tunneling-conductance at $E_F$ at different temperatures.
(a) Along the nearest Fe-Fe direction.
(b) Along the nearest As-As direction.
Solid lines denote results of Gaussian fitting.
(c) Temperature dependence of half-width-at-half-maximum of the fitted Gaussian peak, which represent the vortex-core size.
}
\end{figure*}
We fit each profile with Gaussian function to obtain the half-width-at-half-maximum (HWHM), which can be a good measure of the vortex size.
As shown in Fig.~7(c), HWHM decreases rapidly with decreasing $T$ and tends to saturate below about 2~K.
We note that $T$ dependence of $\xi$ estimated from $B_{c2}$~\cite{Kurita2011JPSJ} is very weak in the $T$ range studied and does not account for the observed strong $T$ dependence.
In addition, HWHM at our lowest $T$=~0.54~K (1.9~nm and 2.5~nm along Fe-Fe and As-As directions, respectively) is much shorter than $\xi(T=0)\sim4.64$~nm.~\cite{Kurita2011JPSJ}
These observations strongly suggest that the KP effect plays an important role in LiFeAs.

The importance of the observation of KP effect in LiFeAs is twofold.
Firstly, to our knowledge, the KP effect has only been detected by muon spin rotation technique.~\cite{Sonier2004JPhys}
Muon spin rotation primarily senses the field or current distribution inside the sample and the vortex-core size is defined as the distance from the vortex center at which the current density reaches the maximum value.~\cite{Sonier2004JPhys}
In contrast to this rather indirect way which requires a specific model, STM/STS gives a direct unambiguous evidence of KP effect.

Secondly, the observation of KP effect and its saturation at low $T$ allows us to estimate the energy of the lowest bound state.
In the quantum-limit vortex, vortex core stops shrinking at $T<\Delta^2/k_BE_F$, because no more low-energy states are available.~\cite{Kramer1974ZPhys,Hayashi1998PRL}
In the presence of impurity scattering, the saturation occurs even at higher $T$.~\cite{Hayashi2005JLTP}
The observed saturation at $T\sim$~2~K sets the upper bound of the lowest bound-state energy associated with the smaller SG to be about 0.2~meV.

\section{Conclusions}

Comprehensive STM/STS experiments on LiFeAs over wide $B$ and $T$ ranges provide new insights into the properties of vortex matter and the electronic states of individual vortices.

Zero-field tunneling spectra contain two superconducting gaps.
By comparing with ARPES results,~\cite{Umezawa2012PRL,Borisenko2012Symmetry,Okazaki_PC} we identify the larger ($\Delta\sim6.0$~meV) and smaller ($\Delta\sim2.9$~meV) gaps as opening on inner and outer hole Fermi surface cylinders, respectively.

Vortices in LiFeAs do not form into a well-ordered lattice while there is a well-developed distance correlation between vortices.
We found that local hexatic correlation retains at low $B$ but the vortex matter contains domain boundaries which consist of alternating vortices with unusual coordination numbers of 5 and 7.
With increasing $B$, the vortex matter approaches to uniformly disordered amorphous state through increasing number of domain boundaries and roughening of each boundary.

LDOS distribution at $E_F$ around a single vortex exhibits tails along the nearest As-As directions and each tail splits in a parallel fashion at elevated energies.
This 4-fold anisotropic shape of the vortex core is associated with the square shape of the outer hole Fermi-surface cylinder.

The multi-gap character of LiFeAs manifests itself in the spatial evolution of LDOS spectrum around a vortex as two quasi-particle branches.
Tunneling spectrum at the vortex center is strongly particle-hole asymmetric, being reminiscent of the behavior expected in the quantum-limit vortex core where the discrete bound states inside the core are smeared out neither by impurity scattering nor thermal broadening.
The quasi-particle branch associated with the larger superconducting gap exhibit step-like energy change as approaching to the vortex center.
This is consistent with the Friedel-like oscillation expected in the quantum-limit vortex.~\cite{Hayashi1998PRL}
However, quasi-particle states associated with the smaller superconducting gap are not compatible with the quantum-limit behavior, although energy-gap like feature is observed even at the vortex center.
Observation of Kramer-Pesch effect, strong $T$ dependence in the vortex-core size, also suggests that energy-separation between the discrete bound states associated with the smaller gap is not large enough to bring about detectable energy gap at the vortex center.
The mechanism which opens the energy gap at the vortex center is an interesting open question.
Nevertheless, we anticipate that observed rich spectroscopic features exposed in the vortex core of LiFeAs serve as a touchstone of various theoretical models for the superconducting state in iron-based materials.

{\it Note added.-} During the completion of the manuscript, we became aware that Chi {\it et al.} also observed two superconducting gaps in LiFeAs by STM/STS~\cite{Chi2012condmat}.

\begin{acknowledgments}
The authors thank M. P. Allan, A. F. Bangura, T. -M. Chuang, J. C. Davis, N. Hayashi, P. J. Hirschfeld, Y. Kato, E. -A. Kim, K. Machida, Y. Nagai, K. Okazaki, A. Rost, M. Takigawa, I. Vekhter, Y. Wang and H. -H. Wen for valuable discussions and comments.
This work has been supported by a Grant-in-Aid for Scientific Research from the Ministry of Education, Culture, Sports, Science and Technology of Japan.
\end{acknowledgments}

\end{document}